\documentclass{article}

\usepackage[final]{tackling_climate_workshop_style}
\usepackage{graphicx}
\usepackage{xcolor}

\usepackage[utf8]{inputenc} 
\usepackage[T1]{fontenc}    
\usepackage{hyperref}       
\usepackage{url}            
\usepackage{booktabs}       
\usepackage{amsfonts}       
\usepackage{nicefrac}       
\usepackage{microtype}      
\usepackage{multirow}
\usepackage{float}
\usepackage{caption}
\usepackage{appendix}
\usepackage{natbib}
\setcitestyle{authoryear}

\title{Enabling Adoption of Regenerative Agriculture through Soil Carbon Copilots}

\author{%
  Margaret Capetz\\
  University of California, Los Angeles\\
  \texttt{mcapetz17@g.ucla.edu} \\
  \And
  Swati Sharma \\
  Microsoft \\
  \texttt{swatisharma@microsoft.com} \\
  \And
  Rafael Padilha \\
  Microsoft Research \\
  \texttt{rpadilha@microsoft.com} \\
  \AND
  Peder Olsen \\
  \texttt{pederao@gmail.com} \\
  \And
  Jessica Wolk \\
   Microsoft \\
   \texttt{jessica.wolk@microsoft.com} \\
   \And
  Emre Kiciman \\
  Microsoft \\
  \texttt{emrek@microsoft.com} \\
  \And
  Ranveer Chandra \\
  Microsoft \\
  \texttt{ranveer@microsoft.com} \\
}

\begin{document}

\maketitle

\begin{abstract}
  Mitigating climate change requires transforming agriculture to minimize environmental impact and build climate resilience. Regenerative agricultural practices enhance soil organic carbon (SOC) levels, thus improving soil health and sequestering carbon. A challenge to increasing regenerative agriculture practices is cheaply measuring SOC over time and understanding how SOC is affected by regenerative agricultural practices and other environmental factors and farm management practices. To address this challenge, we introduce an AI-driven Soil Organic Carbon Copilot that automates the ingestion of complex multi-resolution, multi-modal data to provide large-scale insights into soil health and regenerative practices. Our data includes extreme weather event data (e.g., drought and wildfire incidents), farm management data (e.g., cropland information and tillage predictions), and SOC predictions. We find that integrating public data and specialized models enables large-scale, localized analysis for sustainable agriculture. In comparisons of agricultural practices across California counties, we find evidence that diverse agricultural activity may mitigate the negative effects of tillage; and that while extreme weather conditions heavily affect SOC, composting may mitigate SOC loss. Finally, implementing role-specific personas empowers agronomists, farm consultants, policymakers, and other stakeholders to implement evidence-based strategies that promote sustainable agriculture and build climate resilience. 
\end{abstract}

\vspace{-5pt}
\section{Introduction}
\vspace{-10pt}

The agricultural sector faces the simultaneous challenges of addressing climate change and feeding a growing population. As a result, we must concurrently enhance agricultural productivity, minimize environmental impact, and adapt agriculture to be resilient to a changing climate \cite{WalkeExploringTI}. Previous research demonstrates the potential of regenerative agricultural practices for higher, more nutritious yields, improved soil health, and carbon sequestration capabilities \cite{LAL20041}. Though soil is a complex living medium, we can leverage remote sensing and machine learning to predict SOC, a proxy measurement for soil health \cite{sharma2023knowledgeguidedrepresentationlearning, sharma2023domain}. Soil science research has demonstrated while regenerative practices, including composting and cover cropping, may improve soil, several other external factors, like extreme weather conditions, may lead to SOC loss. 

With the many factors influencing soil health, there is the opportunity to utilize AI-driven data insights to inform sustainable agriculture decisions. Data-driven insights not only have the potential to help farmers directly but also to inform policymakers who develop grants to help fund regenerative agriculture. For example, the California Department of Food and Agriculture has a Healthy Soils Program to fund farms to implement regenerative agriculture practices. It is important to note that there is skepticism about fair access to and effectiveness of funding, among other concerns \cite{farmerperspectives}.

Despite the opportunity of adopting AI for regenerative agriculture, some barriers include speed, scalability, and accessibility. Further, the agriculture industry encompasses a range of stakeholders, including smallholder farmers, large corporations, agronomists, consultants, and policymakers, each with distinct needs and priorities. Stakeholder diversity poses challenges in developing universally effective and flexible solutions. Related work has shown that large language models (LLMs) have the potential to address the challenges mentioned previously, as well as reach various stakeholders~\cite{a2023sustainability, KUSKA2024108924, LLMag}. In this paper, we present a SOC copilot, an AI soil analyst, that enables large-scale analysis of the effects of extreme weather events and land management practices on soil health for a range of agricultural stakeholders to support agriculture and environmental decision-making.

\vspace{-10pt}

\subsection{Contributions}
\vspace{-10pt}

Our AI-enabled SOC Copilot makes significant advancements in data-driven agriculture by enabling serveral new use cases through large-scale data analysis. The copilot automates the complex process of ingesting comprehensive multi-resolution, multi-modal data from public sources, complemented by specialized machine learning models. Our contributions include:
\vspace{-10pt}

\paragraph{Enhanced Large-scale Data Accessibility}
Aggregating and enhancing existing datasets with AI-driven insights enables large-scale soil health analysis. This capability is crucial for advancing regenerative agricultural practices and understanding their impact on SOC. We demonstrate the capabilities of the copilot by generating SOC prediction data for the state of California, USA, for the years 2016 and 2023.

\vspace{-10pt}

\paragraph{Tillage Detection at Scale}
Tillage plays a vital role in influencing changes in SOC. However, obtaining this data is often challenging. We conduct extensive pixel-based tillage detection across California with Coherent Change Detection (CCD) \cite{preiss2006coherent}. To our knowledge, this is the first study to undertake large-scale tillage detection.
\vspace{-10pt}

\paragraph{Personalized Stakeholder Engagement}
We provide tailored personas within the SOC Copilot, enabling customized, data-driven advice for various agricultural stakeholders. This personalized approach ensures that the tool meets the needs of different users, enhancing its utility and effectiveness.
\vspace{-10pt}

\subsection{Opportunity for Impact}
\vspace{-10pt}

The SOC Copilot leverages LLMs to address critical challenges in regenerative agriculture and climate resilience, offering tangible benefits for agricultural practice and policy development.
\vspace{-10pt}

\paragraph{Improved Accessibility and Social Equity}
The SOC Copilot may improve accessibility for diverse stakeholders, including those from economically disadvantaged communities. By offering flexible model configurations, the copilot provides an financially viable solution for farmers from vulnerable communities, such as smallholder farmers. These farmers manage approximately 24\% of world agricultural land \cite{LOWDER201616, RICCIARDI201864} and often lack access to expensive data-driven tools. Our solution ensures that smallholders are included in AI innovation for agriculture.
\vspace{-10pt}

\paragraph{Impactful Insights for Policy and Practice}
Our SOC Copilot provides practical analyses of factors influencing soil health, guiding better agricultural practices and policies. These insights are essential for creating evidence-based policies that enhance climate resilience.
\vspace{-10pt}

\paragraph{Justification and Impact of AI Methods}
Using AI in our solution is essential for managing and analyzing large volumes of complex, multi-modal data. Machine learning techniques provide the necessary scalability and precision to leverage these datasets, offering valuable predictions and insights that would be challenging to obtain through traditional methods.

\section{Methods}
\begin{figure}[h!]
    \centering
    \tiny
        \centering
    \includegraphics[width=0.4\linewidth]{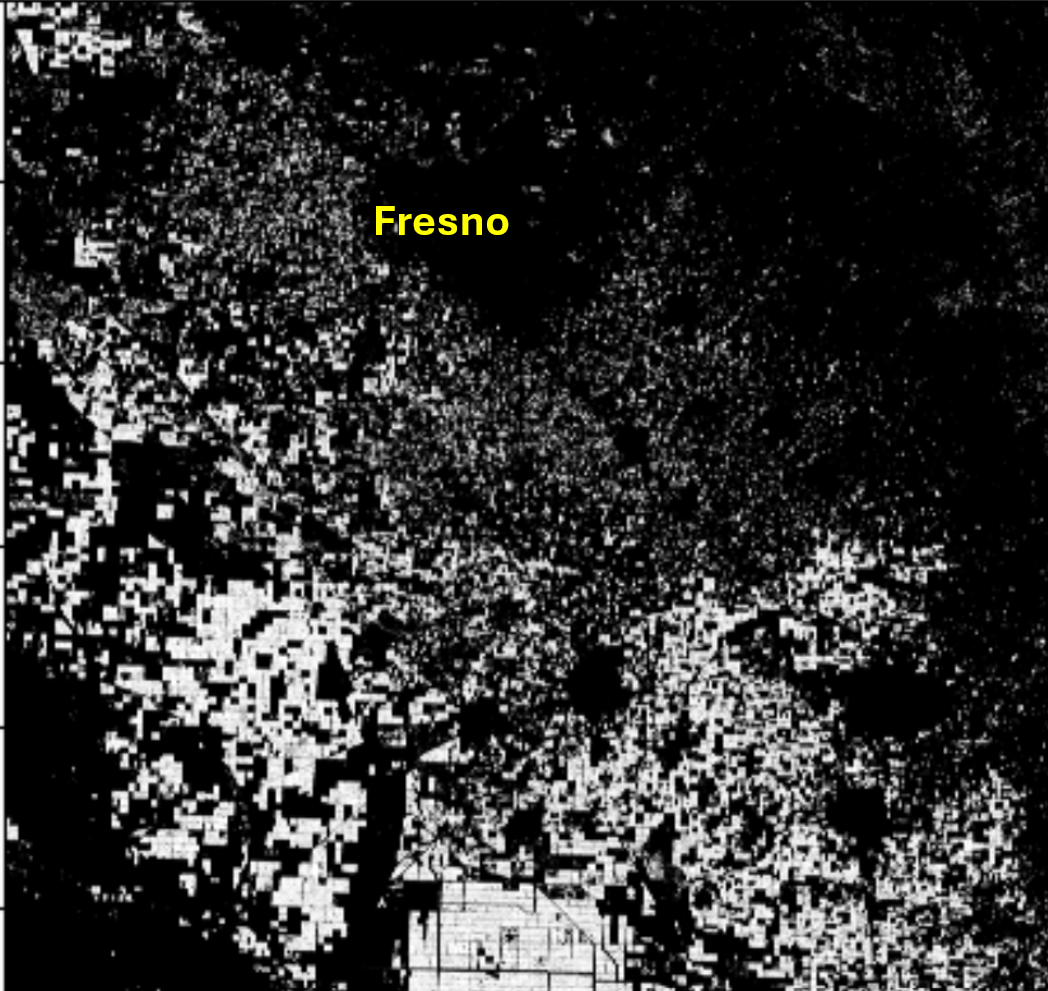}
    \captionof{figure}{Tillage detection for a region of central California for a Sentinel-2 tile (11SKA) which is roughly, 109.8 km by 109.8 km area.}
   \label{fig:ca}
\end{figure}

\paragraph{Data}
\vspace{-10pt}
The SOC copilot ingests both textual and tabular data. First, we provide the copilot with recent soil science publications via unstructured textual data, as shown in Table~\ref{tab:articles}. These articles include domain knowledge in soil science, providing the copilot with insights into how SOC is affected by various environmental and farm management practices. Next, we consider extreme weather event data, including drought data from the US Drought Monitor (USDM) \cite{USDroughtMonitor} and wildfire data from CalFire \cite{calfire}. 
In addition, we consider farm management data including crop type from cropland datalayer (CDL) \cite{CropScape} from the US Department of Agriculture. The copilot also ingests a tillage practice value provided by a specialized tillage detection model based on coherence change detection. Figure~\ref{fig:ca} shows the results of tillage detection. We see the city of Fresno correctly outlined as no-till, and the wilderness in the northeast of the image is also correctly marked as no-till. Furthermore, we can see the outline of individual fields marked as till and the outline of roads cutting through these fields in the lower portion of the image. See Appendix~\ref{appendix:tillage} for further details. We run inference on a SOC prediction model for California (years 2016 and 2023) using Sentinel-2 imagery, location, and Digital Elevation Model (DEM) data and provide tabular inference results to the copilot. See Appendix~\ref{appendix:soc_prediction} for more details.

\paragraph{Copilot Architecture}
\vspace{-10pt}

\begin{figure}[h!]
    \centering
    \tiny
        \centering
   \includegraphics[width=\linewidth]{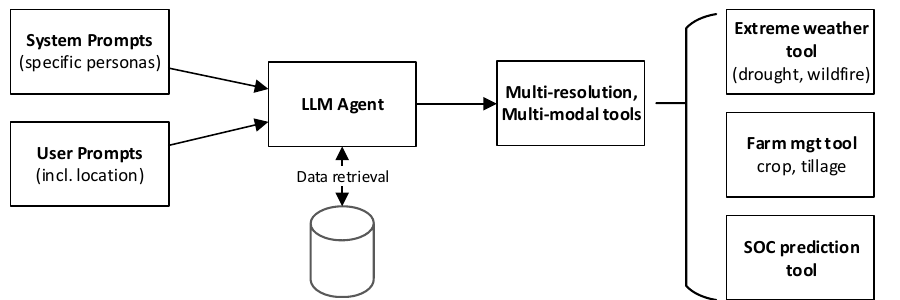}
  \caption{\small SOC Copilot Architecture. The copilot processes user queries containing location data (e.g., county names or coordinates) and customizes responses using role-based personas. It retrieves and analyzes relevant data using multi-resolution, multi-modal tools.}
  \label{fig:architecture}
  \end{figure}

The copilot architecture (Figure~\ref{fig:architecture}) is centered on an LLM Agent that processes queries through a prompt interface. We utilize a GPT-4-Turbo model as the core of our agent, configured via a system prompt to perform specific roles. These roles (Table~\ref{tab:system_prompts}) facilitate personalized stakeholder engagement by providing users with role-specific insights. Upon receiving a prompt, the agent determines which models and data sources are relevant to generate an accurate, comprehensive response. Table~\ref{tab:tools} outlines the essential tools for extracting pertinent data and grounding the copilot's answer with factual and precise information, employing a retrieval augmented generation (RAG) framework. We integrate both structured and unstructured data, including textual scientific literature and tabular environmental and farm management data. The LLM Agent synthesizes the retrieved information from these tools to provide comprehensive answers to the user.

\section{Results and Discussion}
\vspace{-10pt}

We query the SOC copilot with questions regarding how different factors influence SOC change and compared the responses of RAG-enabled SOC and GPT-4-Turbo. We qualitatively show the added value of the RAG-enabled SOC Copilot. 
\vspace{-10pt}
\paragraph{Influence of environmental factors at scale}
We test the large-scale data analysis capabilities of the SOC Copilot vs GPT-4 by asking how extreme weather events impact SOC change and if the trends align with scientific literature. From the question about drought effects in San Joaquin County, we see the SOC Copilot's access to SOC predictions enables it to provide localized, data-driven insights, shown in Table~\ref{tab:soc_gpt4_responses_drought}. We discover that more aggressive soil recovery efforts may be the cause of Sonoma County's SOC gain, while despite less intense wildfire conditions, Merced County experienced SOC loss, shown in Table~\ref{tab:soc_gpt4_responses_wildfire}. RAG enables the SOC Copilot to provide a more nuanced analysis of extreme weather effects on soil health.
\vspace{-10pt}

\paragraph{Effect of adopting regenerative practices}
In addition, we ask farm management-related questions. In Table~\ref{tab:soc_gpt4_responses_tillage}, we compare the effects of the varied tillage practices of Monterey and Tulare County. We find that "despite high-intensity tillage,
Tulare has managed to maintain relatively high SOC levels, likely through integrated soil health practices and diverse
cropping systems contributing positively to soil organic matter. Conversely, despite no-tillage, the SOC in Monterey
has decreased, likely influenced by severe external stresses like drought impacting biomass production negatively." Also, we obtain a list of farms from the California Healthy Soils Program, specifically farms that applied to receive funding from Incentive Grants \cite{HealthySoilsProgram}. With information including funding status and regenerative practices, we ask farm-specific questions and compare regenerative agricultural practices in Table~\ref{tab:soc_gpt4_responses_practice_tiny}. We learn the effectiveness of regenerative agriculture practices is heavily moderated by external pressures like extreme weather events, which may call for more holistic and aggressive regenerative strategies. In this comparison, the SOC Copilot suggests compost is effective in protecting against SOC loss amidst extreme environmental conditions, which aligns with \cite{CompostCoverCrop}.

\begin{figure}[h!]
        \centering
\begin{tabular}{p{2cm} p{10cm}}  
\multicolumn{2}{c}{\textbf{Response Excerpts of Effect of Farm Management on SOC}} \\ \hline
\multicolumn{2}{p{12cm}}{\textbf{Question:} "Gaviota Givings, Orella Ranch in Santa Barbara County implemented the regenerative practice of planting in 2021-22. How does this influence the change in SOC among other factors like drought and wildfire? Based on your role, what specific insights can you draw?"} \\ \hline
\textbf{Model} & \textbf{Response Excerpt} \\ \hline
\textbf{SOC Copilot} & 
[...] \textcolor{blue}{SOC Changes:
                - **Riverside County**:
                  - SOC 2016: 2.99\%
                  - SOC 2023: 0.94\%}
                  - A significant reduction in SOC despite regenerative planting, possibly due to other factors like drought and frequent wildfires which can contribute to organic matter loss.
                - \textcolor{blue}{**Marin County**:
                  - SOC 2016: 1.96\%
                  - SOC 2023: 1.92\%}
                  - Slightly decreased SOC, which might suggest that composting helped in buffering the SOC loss considering the drought and wildfire pressures. [...]
                  \textcolor{blue}{The results suggest that while regenerative practices like planting and composting can support soil carbon sequestration, their effectiveness can be heavily moderated by external pressures such as drought and wildfires.} This is evident from the stark difference in SOC trends between Riverside and Marin given similar drought histories but different intensities and frequencies of wildfires and regenerative strategies.[...]\\ \hline
\textbf{GPT-4} & 
[...] Conclusion
                Regenerative practices like those implemented by Gable Farms and Doug and Cathy Ielmorini Dairy are generally beneficial for increasing SOC levels, which can help improve soil health, enhance productivity, and mitigate climate change. \textcolor{blue}{However, the effectiveness of these practices can be moderated by external factors such as drought and wildfire, highlighting the need for adaptive management strategies in the face of environmental changes.} \\ \hline
\end{tabular}
\caption{Comparison of SOC Copilot and GPT-4 Responses. For a query comparing regenerative practices' impact on SOC in Riverside vs Marin counties, GPT-4 offers general analysis while SOC Copilot provides nuanced data-driven insights, highlighting environmental conditions as the primary driver of SOC change. See Table~\ref{tab:soc_gpt4_responses_practice} for complete responses.}
\label{tab:soc_gpt4_responses_practice_tiny}
\end{figure}

\paragraph{Personalized Stakeholder Engagement}
We also experiment with various copilot personas listed in Table~\ref{tab:system_prompts}. The results in Table~\ref{tab:persona_responses_excerpts} demonstrate the SOC Copilot's ability to reach numerous stakeholders and provide actionable, data-driven recommendations. Despite the range of stakeholders in the agriculture industry, the SOC Copilot can address stakeholders' distinct priorities. The agronomist persona provides soil-specific insights, the farm consultant provides higher-level advice about management practices, while the policymaker persona shares policy-level recommendations, such as "foster research initiatives that systematically analyze the long-term impacts of regenerative practices on SOC under varying climatic stress conditions to refine agricultural policy making and support."

\begin{figure}[h!]
        \centering
\begin{tabular}{p{3cm} p{9cm}}  
\multicolumn{2}{c}{\textbf{Highlighted Excerpts of Personalized Stakeholder Responses}} \\ \hline
\multicolumn{2}{p{12cm}}{\textbf{Question:} "Gaviota Givings, Orella Ranch in Santa Barbara County implemented the regenerative practice of planting in 2021-22. How does this influence the change in SOC among other factors like drought and wildfire? Based on your role, what specific insights can you draw?"} \\ \hline
\textbf{Role} & \textbf{Response Excerpt} \\ \hline
\textbf{Agronomist} & 
"Recommendations: \textcolor{blue}{1. **Enhanced Water Management:** Implementing more efficient irrigation systems and drought-tolerant crop varieties could mitigate some of the drought-related declines in crop productivity and, indirectly, SOC levels."}
\\ \hline
\textbf{Farm Consultant} & 
"Insights and Advice:
\textcolor{blue}{3. **Integrate Additional Resilience Practices:** Given the recurrent wildfires, integrating fire resilience strategies, such as creating firebreaks, choosing fire-resistant plant species, or even controlled burns under safe conditions, could help protect the soil and vegetation."} \\ \hline
\textbf{Policymaker} & 
"Recommendations for Policy and Further Actions:
\textcolor{blue}{**Research and Data Analysis**: Foster research initiatives that systematically analyze the long-term impacts of regenerative practices on SOC under varying climatic stress conditions to refine agricultural policy making and support.} \\ \hline
\end{tabular}
\caption{Excerpts of stakeholder-specific SOC Copilot responses. The Agronomist, Farm Consultant, and Policy Maker roles provide role-specific insights for regenerative practices' SOC impact at Orella Ranch, Santa Barbara County. See Table~\ref{tab:persona_responses} for expanded responses.}
\label{tab:persona_responses_excerpts}
\end{figure}




\section{Conclusion}
\vspace{-10pt}

Our SOC Copilot advances large-scale analysis of the factors influencing soil health, providing a rapid, scalable, and accessible solution for AI-driven regenerative agriculture. By leveraging AI to understand the complex interplay of variables affecting soil organic carbon (SOC), we demonstrate its potential to guide the development of more effective regenerative agricultural practices and inform related policy decisions by providing personalized stakeholder engagement. We also discover nuanced insights, including how diverse agricultural activity may mitigate adverse effects of tillage, extreme weather events may influence SOC more than regenerative practices, and composting may protect against SOC loss. In future work, we will consider quantitative evaluation metrics. As we move toward a future of sustainable agriculture, these AI-driven insights can play a crucial role in enhancing soil carbon sequestration, building climate resilience, and ultimately mitigating climate change.


\bibliographystyle{plainnat}
\bibliography{main}

\begin{thebibliography}{30}
\providecommand{\natexlab}[1]{#1}
\providecommand{\url}[1]{\texttt{#1}}
\expandafter\ifx\csname urlstyle\endcsname\relax
  \providecommand{\doi}[1]{doi: #1}\else
  \providecommand{\doi}{doi: \begingroup \urlstyle{rm}\Url}\fi

\bibitem[pat(2021)]{patel}
Soil texture and environmental conditions influence the biogeochemical responses of soils to drought and flooding.
\newblock 2021.
\newblock URL \url{https://doi.org/10.1038/s43247-021-00198-4}.

\bibitem[Sal(2024)]{Salgado}
Impact of wildfire recurrence on soil properties and organic carbon fractions.
\newblock 2024.
\newblock \doi{10.1016/j.jenvman.2024.120293}.

\bibitem[A et~al.(2023)A, V, Raja, and V]{a2023sustainability}
Rajagopal A, Nirmala V, Immanuel Raja, and Arun V.
\newblock Sustainability ai copilot: Analyze \& ideate at scale to enable positive impact.
\newblock In \emph{NeurIPS 2023 Workshop on Tackling Climate Change with Machine Learning}, 2023.
\newblock URL \url{https://www.climatechange.ai/papers/neurips2023/118}.

\bibitem[{Barbato, C.T., Strong, A.L.}(2023)]{farmerperspectives}
{Barbato, C.T., Strong, A.L.}
\newblock {Farmer perspectives on carbon markets incentivizing agricultural soil carbon sequestration}, 2023.
\newblock URL \url{https://doi.org/10.1038/s44168-023-00055-4}.

\bibitem[Blanco-Canqui and Lal(2008)]{Blanco-Canqui}
Humberto Blanco-Canqui and R.~Lal.
\newblock No-tillage and soil-profile carbon sequestration: An on-farm assessment.
\newblock \emph{Soil Science Society of America Journal}, 72\penalty0 (3):\penalty0 693--701, 2008.
\newblock \doi{https://doi.org/10.2136/sssaj2007.0233}.
\newblock URL \url{https://acsess.onlinelibrary.wiley.com/doi/abs/10.2136/sssaj2007.0233}.

\bibitem[Boryan et~al.(2011)Boryan, Yang, Mueller, and Craig]{boryan2011monitoring}
Claire Boryan, Zhengwei Yang, Rick Mueller, and Mike Craig.
\newblock Monitoring us agriculture: the us department of agriculture, national agricultural statistics service, cropland data layer program.
\newblock \emph{Geocarto International}, 26\penalty0 (5):\penalty0 341--358, 2011.

\bibitem[{California Department of Food and Agriculture}(2024)]{HealthySoilsProgram}
{California Department of Food and Agriculture}.
\newblock Healthy soils program - incentives program.
\newblock \url{https://www.cdfa.ca.gov/oefi/healthysoils/IncentivesProgram.html}, 2024.
\newblock Accessed: 2024-08-16.

\bibitem[{California Department of Forestry and Fire Protection}(2024)]{calfire}
{California Department of Forestry and Fire Protection}.
\newblock Fire incidents.
\newblock \url{https://www.fire.ca.gov/incidents}, 2024.
\newblock Accessed: 2024-08-22.

\bibitem[et~al(2019)]{CompostCoverCrop}
Tautges et~al.
\newblock Deep soil inventories reveal that impacts of cover crops and compost on soil carbon sequestration differ in surface and subsurface soils.
\newblock \emph{Global Change Biology}, 2019.
\newblock URL \url{https://doi.org/10.1111/gcb.14762}.

\bibitem[Ghimire et~al.(2019)Ghimire, Bista, and Machado]{Ghimire}
R.~Ghimire, P.~Bista, and S.~Machado.
\newblock Long-term management effects and temperature sensitivity of soil organic carbon in grassland and agricultural soils.
\newblock 2019.
\newblock \doi{https://doi.org/10.1016/j.agee.2024.108925}.
\newblock URL \url{https://doi.org/10.1038/s41598-019-48237-7}.

\bibitem[Kim et~al.(2022)Kim, Daly, Flesch, Coates, and Hernandez-Ramirez]{KIM2022108805}
Keunbae Kim, Erin~J. Daly, Thomas~K. Flesch, Trevor~W. Coates, and Guillermo Hernandez-Ramirez.
\newblock Carbon and water dynamics of a perennial versus an annual grain crop in temperate agroecosystems.
\newblock \emph{Agricultural and Forest Meteorology}, 314:\penalty0 108805, 2022.
\newblock ISSN 0168-1923.
\newblock \doi{https://doi.org/10.1016/j.agrformet.2021.108805}.
\newblock URL \url{https://www.sciencedirect.com/science/article/pii/S0168192321004913}.

\bibitem[Kuska et~al.(2024)Kuska, Wahabzada, and Paulus]{KUSKA2024108924}
Matheus~Thomas Kuska, Mirwaes Wahabzada, and Stefan Paulus.
\newblock Ai for crop production – where can large language models (llms) provide substantial value?
\newblock \emph{Computers and Electronics in Agriculture}, 221:\penalty0 108924, 2024.
\newblock ISSN 0168-1699.
\newblock \doi{https://doi.org/10.1016/j.compag.2024.108924}.
\newblock URL \url{https://www.sciencedirect.com/science/article/pii/S0168169924003156}.

\bibitem[Lal(2004)]{LAL20041}
R.~Lal.
\newblock Soil carbon sequestration to mitigate climate change.
\newblock \emph{Geoderma}, 123\penalty0 (1):\penalty0 1--22, 2004.
\newblock ISSN 0016-7061.
\newblock \doi{https://doi.org/10.1016/j.geoderma.2004.01.032}.
\newblock URL \url{https://www.sciencedirect.com/science/article/pii/S0016706104000266}.

\bibitem[Li et~al.(2023)Li, Jin, He, Wang, Sun, Luo, Huang, Li, Chang, Wang, and Wei]{LI2023106797}
Xiaoying Li, Huijun Jin, Ruixia He, Hongwei Wang, Long Sun, Dongliang Luo, Yadong Huang, Yan Li, Xiaoli Chang, Lizhong Wang, and Changlei Wei.
\newblock Impact of wildfire on soil carbon and nitrogen storage and vegetation succession in the nanweng'he national natural wetlands reserve, northeast china.
\newblock \emph{CATENA}, 221:\penalty0 106797, 2023.
\newblock ISSN 0341-8162.
\newblock \doi{https://doi.org/10.1016/j.catena.2022.106797}.
\newblock URL \url{https://www.sciencedirect.com/science/article/pii/S0341816222007834}.

\bibitem[Lowder et~al.(2016)Lowder, Skoet, and Raney]{LOWDER201616}
Sarah~K. Lowder, Jakob Skoet, and Terri Raney.
\newblock The number, size, and distribution of farms, smallholder farms, and family farms worldwide.
\newblock \emph{World Development}, 87:\penalty0 16--29, 2016.
\newblock ISSN 0305-750X.
\newblock \doi{https://doi.org/10.1016/j.worlddev.2015.10.041}.
\newblock URL \url{https://www.sciencedirect.com/science/article/pii/S0305750X15002703}.

\bibitem[{Microsoft Planetary Computer}(n.d.)]{sentinel2_l2a}
{Microsoft Planetary Computer}.
\newblock Sentinel-2 l2a.
\newblock \url{https://planetarycomputer.microsoft.com/dataset/sentinel-2-l2a}, n.d.
\newblock Accessed: 2024-08-29.

\bibitem[{National Drought Mitigation Center, University of Nebraska-Lincoln}(2024)]{USDroughtMonitor}
{National Drought Mitigation Center, University of Nebraska-Lincoln}.
\newblock U.s. drought monitor.
\newblock \url{https://droughtmonitor.unl.edu/CurrentMap.aspx}, 2024.
\newblock Accessed: 2024-08-16.

\bibitem[Preiss and Stacy(2006)]{preiss2006coherent}
Mark Preiss and Nicholas~JS Stacy.
\newblock Coherent change detection: Theoretical description and experimental results.
\newblock Technical report, 2006.

\bibitem[Ricciardi et~al.(2018)Ricciardi, Ramankutty, Mehrabi, Jarvis, and Chookolingo]{RICCIARDI201864}
Vincent Ricciardi, Navin Ramankutty, Zia Mehrabi, Larissa Jarvis, and Brenton Chookolingo.
\newblock How much of the world's food do smallholders produce?
\newblock \emph{Global Food Security}, 17:\penalty0 64--72, 2018.
\newblock ISSN 2211-9124.
\newblock \doi{https://doi.org/10.1016/j.gfs.2018.05.002}.
\newblock URL \url{https://www.sciencedirect.com/science/article/pii/S2211912417301293}.

\bibitem[Robertson et~al.(2023)Robertson, McNairn, van~der Kooij, Jiao, Ihuoma, and Joosse]{robertson2023monitoring}
Laura~Dingle Robertson, Heather McNairn, Marco van~der Kooij, Xianfeng Jiao, Samuel Ihuoma, and Pamela Joosse.
\newblock Monitoring autumn agriculture activities using synthetic aperture radar (sar) and coherence change detection.
\newblock \emph{Heliyon}, 9\penalty0 (6), 2023.

\bibitem[Sandwell et~al.(2011)Sandwell, Mellors, Tong, Wei, and Wessel]{sandwell2011gmtsar}
David Sandwell, Rob Mellors, Xiaopeng Tong, Matt Wei, and Paul Wessel.
\newblock Gmtsar: An insar processing system based on generic mapping tools.
\newblock 2011.

\bibitem[Satalino et~al.(2018)Satalino, Mattia, Balenzano, Lovergine, Rinaldi, De~Santis, Ruggieri, Garc{\'\i}a, G{\'o}mez, Ceschia, et~al.]{satalino2018sentinel}
Giuseppe Satalino, Francesco Mattia, Anna Balenzano, Francesco~P Lovergine, Michele Rinaldi, Angelo~Pio De~Santis, Sergio Ruggieri, DA~Nafr{\'\i}a Garc{\'\i}a, V~Paredes G{\'o}mez, Eric Ceschia, et~al.
\newblock Sentinel-1 \& sentinel-2 data for soil tillage change detection.
\newblock In \emph{IGARSS 2018-2018 IEEE International Geoscience and Remote Sensing Symposium}, pages 6627--6630. IEEE, 2018.

\bibitem[Shang et~al.(2024)Shang, Olesen, Lærke, Manevski, and Chen]{SHANG2024108925}
Yiwei Shang, Jørgen~Eivind Olesen, Poul~Erik Lærke, Kiril Manevski, and Ji~Chen.
\newblock Perennial cropping systems increased topsoil carbon and nitrogen stocks over annual systems—a nine-year field study.
\newblock \emph{Agriculture, Ecosystems \& Environment}, 365:\penalty0 108925, 2024.
\newblock ISSN 0167-8809.
\newblock \doi{https://doi.org/10.1016/j.agee.2024.108925}.
\newblock URL \url{https://www.sciencedirect.com/science/article/pii/S0167880924000434}.

\bibitem[Sharma et~al.(2023{\natexlab{a}})Sharma, Sharma, Liu, Tushir, Neal, Ness, Crawford, Kiciman, and Chandra]{sharma2023knowledgeguidedrepresentationlearning}
Somya Sharma, Swati Sharma, Licheng Liu, Rishabh Tushir, Andy Neal, Robert Ness, John Crawford, Emre Kiciman, and Ranveer Chandra.
\newblock Knowledge guided representation learning and causal structure learning in soil science, 2023{\natexlab{a}}.
\newblock URL \url{https://arxiv.org/abs/2306.09302}.

\bibitem[Sharma et~al.(2023{\natexlab{b}})Sharma, Sharma, Padilha, Kiciman, and Chandra]{sharma2023domain}
Somya Sharma, Swati Sharma, Rafael Padilha, Emre Kiciman, and Ranveer Chandra.
\newblock Domain adaptation for sustainable soil management using causal and contrastive constraint minimization.
\newblock In \emph{NeurIPS 2023 Workshop on Tackling Climate Change with Machine Learning}, 2023{\natexlab{b}}.
\newblock URL \url{https://www.climatechange.ai/papers/neurips2023/107}.

\bibitem[Soares et~al.(2023)Soares, Harrison, Kalantari, Zhao, and Ferreira]{Soares_2023}
Pedro~R Soares, Matthew~T Harrison, Zahra Kalantari, Wenwu Zhao, and Carla S~S Ferreira.
\newblock Drought effects on soil organic carbon under different agricultural systems.
\newblock \emph{Environmental Research Communications}, 5\penalty0 (11):\penalty0 112001, nov 2023.
\newblock \doi{10.1088/2515-7620/ad04f5}.
\newblock URL \url{https://dx.doi.org/10.1088/2515-7620/ad04f5}.

\bibitem[Tzachor et~al.(2023)Tzachor, Devare, Richards, Pypers, Ghosh, Koo, Johal, and King]{LLMag}
A.~Tzachor, M.~Devare, C.~Richards, P.~Pypers, A.~Ghosh, J.~Koo, S.~Johal, and B.~King.
\newblock Large language models and agricultural extension services.
\newblock \emph{Nature Food}, 4\penalty0 (11):\penalty0 941--948, 2023.
\newblock \doi{10.1038/s43016-023-00867-x}.
\newblock URL \url{https://doi.org/10.1038/s43016-023-00867-x}.

\bibitem[{U.S. Department of Agriculture, National Agricultural Statistics Service}(2024)]{CropScape}
{U.S. Department of Agriculture, National Agricultural Statistics Service}.
\newblock Cropscape - cropland data layer.
\newblock \url{https://catalog.data.gov/dataset/cropscape-cropland-data-layer}, 2024.
\newblock Accessed: 2024-08-16.

\bibitem[Walke et~al.(2022)Walke, P.K.Srivastava, and Sharma]{WalkeExploringTI}
Anil Walke, P.K.Srivastava, and Deepak Sharma.
\newblock Exploring the impact of artificial intelligence on sustainable agriculture: A case study of precision farming.
\newblock 2022.
\newblock URL \url{https://api.semanticscholar.org/CorpusID:264351864}.

\bibitem[Xu et~al.(2022)Xu, Amgain, Rabbany, Capasso, Korus, Swanson, and Bhadha]{Xu}
Nan Xu, Naba~R. Amgain, Abul Rabbany, Jay Capasso, Kevin Korus, Stewart Swanson, and Jehangir~H. Bhadha.
\newblock Interaction of soil health indicators to different regenerative farming practices on mineral soils.
\newblock \emph{Agrosystems, Geosciences \& Environment}, 5\penalty0 (1):\penalty0 e20243, 2022.
\newblock \doi{https://doi.org/10.1002/agg2.20243}.
\newblock URL \url{https://acsess.onlinelibrary.wiley.com/doi/abs/10.1002/agg2.20243}.

\end{thebibliography}

\appendix
\section{Input Data Sources}
\subsection{SOC prediction}
\label{appendix:soc_prediction}
We run inference on a SOC prediction model \cite{sharma2023domain} for the state of California for the years 2016 and 2023 at a resolution of 50m. The model's inputs include satellite imagery data from Sentinel-2, location information (latitude and longitude) and topology (DEM).  We download sentinel 2 imagery for March to August months for both years. We remove cloudy images using the scene classification layer (SCL) available with Sentinel 2 \cite{sentinel2_l2a}. The SOC model outputs a SOC value for each  5 pixel by 5 pixel area for each Sentinel 2 image. The SOC predictions for each image are then averaged to produce the yearly pixel by pixel prediction. After processing the raw SOC raster data, we provide tabular data to the copilot. 

\subsection{Scientific literature}
Table ~\ref{tab:articles} shows a list of recent articles we added to the copilot to get context on wildfire occurrences, drought occurrence and effect of management practices. 
 \begin{table}[h!]
\centering
\begin{tabular}{p{3cm} p{11cm}}
\hline
\textbf{Topic} & \textbf{Article} \\ \hline
\small
\multirow{2}{3cm}{Drought} 
& Drought effects on soil organic carbon under different agricultural systems \cite{Soares_2023} \\ \cline{2-2}
& Soil texture and environmental conditions influence the biogeochemical responses of soils to drought and flooding, 2022, Communications Earth \& Environment (nature.com) \cite{patel} \\ \hline

\multirow{2}{3cm}{Wildfire} 
& Impact of wildfire recurrence on soil properties and organic carbon fractions, PubMed (nih.gov) \cite{Salgado} \\ \cline{2-2}
& Impact of wildfire on soil carbon and nitrogen storage and vegetation succession in the Nanweng'he National Natural Wetlands Reserve, Northeast China, ScienceDirect \cite{LI2023106797} \\ \hline

\multirow{2}{3cm}{Crop} 
& Carbon and water dynamics of a perennial versus an annual grain crop in temperate agroecosystems, ScienceDirect \cite{KIM2022108805} \\ \cline{2-2}
& Perennial cropping systems increased topsoil carbon and nitrogen stocks over annual systems—a nine-year field study, ScienceDirect \cite{SHANG2024108925} \\ \hline

\multirow{3}{3cm}{Practices} 
& Long-term Management Effects and Temperature Sensitivity of Soil Organic Carbon in Grassland and Agricultural Soils, 2022, Scientific Reports (nature.com) \cite{Ghimire} \\ \cline{2-2}
& Interaction of soil health indicators to different regenerative farming practices on mineral soils, 2022, Agrosystems, Geosciences \& Environment (Wiley Online Library) \cite{Xu} \\ \cline{2-2}
& No-Tillage and Soil-Profile Carbon Sequestration: An On-Farm Assessment, 2008, Soil Science Society of America \cite{Blanco-Canqui} \\ \hline

\end{tabular}
\caption{Recent hand-selected articles related to various topics. Providing scientific literature to the copilot enables responses to be grounded in recent research. The hand-selection of articles provides another layer of customization for the SOC Copilot.}
\label{tab:articles}
\end{table}

\subsection{Tillage Detection}
\label{appendix:tillage}
\subsubsection{Coherent Change Detection}

Coherent Change Detection (CCD) uses two time-lapsed satellite synthetic aperture radar (SAR) images of the same geographic area to detect and measure changes to that area, \cite{preiss2006coherent}.  The coherence between two zero-mean complex SAR signals $s_{t_1}$, $s_{t_2}$ is defined to be
\begin{equation}\label{eq:coherence}
\Delta_{t_1t_2}e^{i\Phi_{t_1t_2}} = \frac{E[s_{t_1} s_{t_2}^*]}{\sqrt{E[|s_{t_1}|^2]E[|s_{t_2}|^2]}},
\end{equation}
where $\Phi_{t_1t_2}$ is the interferometric phase and $\Delta_{t_1t_2}$ the coherence.  The interferometric phase $\Phi_{t_1t_2}$, is determined by the baseline offset between the primary and repeat pass collections, the terrain topography and any displacement of the scattering scene between the two data collections.  For the case of random, zero mean scatterer displacement the displacement is zero and change detection can be performed on the associated loss in the magnitude of the cross correlation coefficient, $\Delta_{t_1t_2}$.  The coherence can detect changes as small as 1/4 of the wavelength in the satellite's line-of-sight (LOS).  Sentinel-1 is a SAR C-band satellite with a wavelength of 5.6cm which means changes in the vertical direction that are almost as small as 1cm can be detected. One instance of such a change is a tillage event in agriculture and it is well known that tilling can be detected using CCD, \cite{satalino2018sentinel}.  Other management practices such as harvesting are also detected by CCD, \cite{robertson2023monitoring}.  It should be noted that when the intra-satellite distance (the baseline) is large, the difference in the imaging geometry itself results in a loss of coherence.  Thus, only satellite pairs with a small baseline should be considered (we used $\leq$100m). Moreover, for the expected values in Equation \ref{eq:coherence} we averaged all SAR observations in a $50m\times 50m$ region for all Sentinel-1 SAR $5m\times 20m$ observations in the region.  We used the GMTSAR library to compute the coherence, \cite{sandwell2011gmtsar}.  

\begin{figure}
  \centering
  \includegraphics[width=1.0\linewidth]{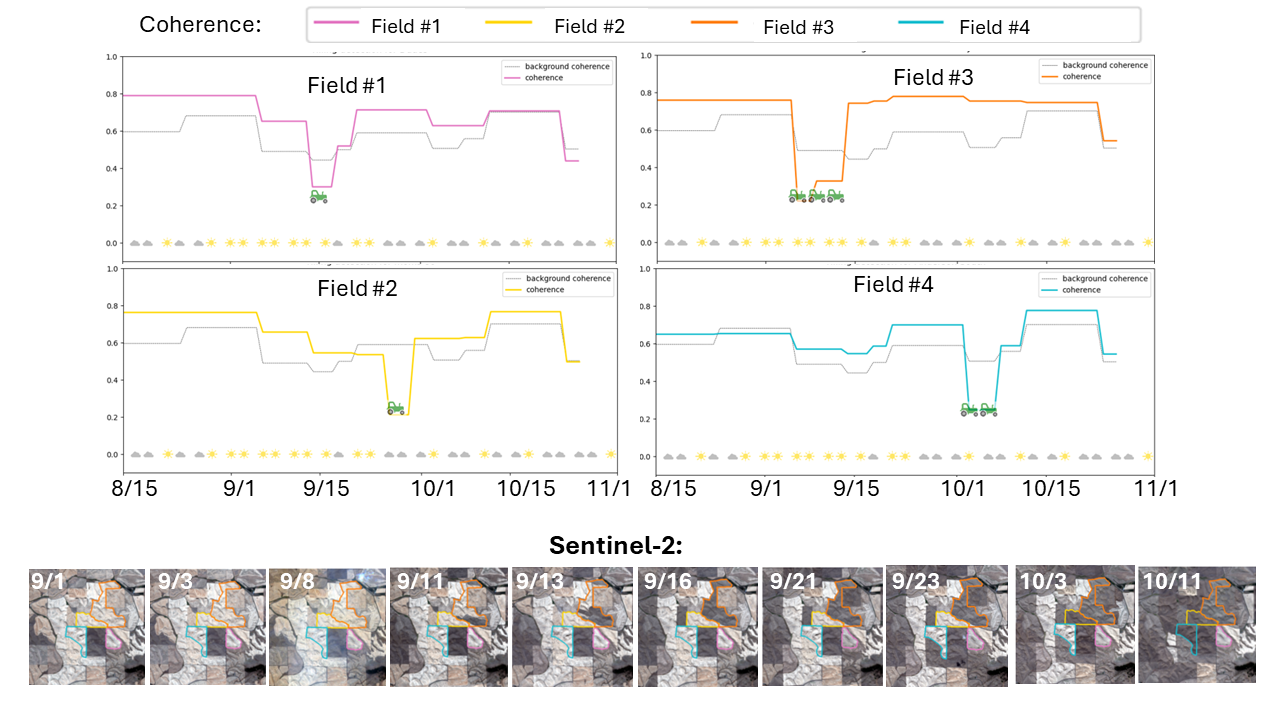}
  \caption{Tillage detection for 4 fields in Washington state.}\label{fig:ccd}
\end{figure}

\subsubsection{Detecting tillage using CCD}
In Figure \ref{fig:ccd} we show detection of tillage for winter wheat fields in the state of Washington where the tillage date was known.  All cloud free Sentinel-2 images in the relevant time window are also shown so the reader can verify the claims.  The farmer reported tillage dates of 9/15, 9/14 and 10/5 for fields 2-4, while field 1 was no-till.  The Figure shows a (weak) false detection for field 1, a slightly wrong date for field 2, tilling for two consecutive pairs of satellite passes for field 3 and a correct detection for field 4.  After checking with the farmer, the date detected for field 2 which is somewhat later, appears to be correct. The recorded tillage date was in this case the intended date.  Field 3 was very large and tilling persisted over several days, but only the start date was recorded.  We can see the partial tilling both in the optical and the coherence images.  We do not know the reason for the false detection for field 1, but it appears not to be weather related.  The interested reader can zoom into the optical imagery in Figure \ref{fig:ccd} and independently verify these claims.

\subsubsection{Pixel based tillage detection in California}
The tilling time and crop type will not in general be known as it was in the case of Winter Wheat in Washington.  Since CCD will detect any kind of change it then becomes necessary to reduce the number of false positives.  If there is any crop present its growth will register as change.  These changes can be eliminated by detecting presence of bare soil (and absence of vegetation).  We ran experiments for a 100km $\times$ 100km region in California corresponding to the Sentinel-2 tile \texttt{11SKA} where multi-cropping is common.  When several crops are grown on the same field it becomes difficult to assess the time of tilling.  We utilized a conservative bare soil detector, namely the Bare Soil Index (BSI) defined by
$$
\mathtt{BSI} = \frac{\mathrm{SWIR}_1+\mathrm{blue}-(\mathrm{red}+\mathrm{NIR})}{\mathrm{SWIR}_1+\mathrm{blue}+\mathrm{red}+\mathrm{NIR}}
$$
where NIR is the near infrared band and $\mathrm{SWIR}_1$ is the short wave infrared band known as \texttt{B11} for the Sentinel-2 satellite.  Our conservative tillage detector uses $\mathtt{BSI}>0.06$ as a bare soil detector, and tillage is defined as a local change during the time soil is bare.  Cases where every time window is detected as change are presumed to be false positives and tilling areas with a width or height less than 3 pixels where assumed to be roads and removed.  Figure~\ref{fig:ca} shows the tillage prediction for each pixel for the year 2019.

\begin{table}[h!]
\centering
\small
\begin{tabular}{l|cc}
{Crop type} & {Till} & {no till}\\\hline
Grass/Pasture            &  1.36 & 98.64\\
Almonds                  &  6.26 & 93.74\\
Grapes                   & 23.55 & 76.45\\
Fallow/Idle Cropland     &  4.27 & 95.73\\
Cotton                   & 75.77 & 24.23\\
Pistachios               &  6.06 & 93.94\\
Alfalfa                  & 32.56 & 67.44\\
Developed/Open Space     &  6.65 & 93.35\\
Shrubland                &  2.18 & 97.82\\
Citrus                   &  2.46 & 97.54\\
Walnuts                  &  9.92 & 90.08\\
Winter Wheat             & 22.78 & 77.22\\
Developed/Med Intensity  &  0.46 & 99.54\\
Tomatoes                 & 78.41 & 21.59\\
Dbl Crop WinWht/Corn     & 75.99 & 24.01\\
Evergreen Forest         &  1.11 & 98.89\\
Developed/Low Intensity  &  2.52 & 97.48\\
Other Tree Crops         &  7.34 & 92.66\\
Corn                     & 71.56 & 28.44\\
Dbl Crop Triticale/Corn  & 78.39 & 21.61\\
\end{tabular}
\bigskip
\caption{Tilling predictions for Sentinel-2 tile \texttt{11SKA} in California}\label{tab:ca}
\end{table}

Using the crops detected from Cropland Data Layer, \cite{boryan2011monitoring} we found the breakdown of till versus no-till in Table~\ref{tab:ca} for the crops grown in the region.  Although we do not have ground truth data for the pixels in the region, the till versus no-till breakdown by crop type appears to be reasonable on the whole and in Figure~\ref{fig:ca} we see the city of Fresno correctly outlined as no-till and the wilderness in the northeast of the image is also correctly marked as no-till.  Furthermore we can see the outline of individual fields that are marked as till as well as the outline of roads cutting through these fields in the lower portion of the image.  This large region of tilled fields had a dominant crop of Cotton that is known to be commonly tilled.

\section{Copilot Architecture Details}

\begin{table}[h]
\centering
\small
\begin{tabular}{|l|p{12cm}|}
\hline
\textbf{Tool} & \textbf{Description} \\ \hline
\texttt{SOC Prediction} & Always use this tool to get the soil organic carbon prediction in 2016 and 2023 in a county. \\ \hline
\texttt{Drought Conditions} & Provides the drought conditions of a county. \\ \hline
\texttt{Wildfire Incidents} & Provides the wildfire incidents that occurred in a county. \\ \hline
\texttt{Crop Types and Years} & Provides the crop types and corresponding years in a county. \\ \hline
\texttt{Tillage Scale} & Provides the tillage in 2019 in a county on a scale of 0 to 1 where 0 is no \\  & till, 1 is conventional tilling. \\ \hline
\texttt{Support Arguments} & Use this tool to find arguments to support your hypothesis with topics related to wildfire, drought, agricultural practices, or crops. \\ \hline
\end{tabular}
\caption{Tool descriptions. Within the copilot architecture, the LLM agent has access to a suite of tools for data extraction. The LLM agent is provided with tools for each data source.}
\label{tab:tools}
\end{table}

\begin{table}[h!]
\centering
\begin{tabular}{p{3cm} p{10cm}}
\multicolumn{2}{c}{\textbf{SOC Copilot Stakeholder Personas}} \\ \hline
\textbf{Role} & \textbf{System Prompt} \\ \hline
\textbf{Agronomist} & 
You are an AI assistant specialized in agronomy, with a focus on soil health and crop productivity. You have access to tools that analyze soil organic carbon (SOC) as a proxy for soil health, along with county-level data on drought, wildfire, and cropland conditions. You provide insights and recommendations for optimizing soil management and enhancing agricultural sustainability. \\ \hline
\textbf{Farm Consultant} & 
You are an AI assistant acting as a farm consultant, helping farmers make informed decisions about land management. You have access to tools that assess soil organic carbon (SOC) as a key indicator of soil health, alongside data on drought, wildfire, and cropland conditions at the county level. You offer actionable advice to improve farm productivity and resilience. \\ \hline
\textbf{Policymaker} & 
You are an AI assistant supporting policy makers in developing and implementing agricultural policies. You use tools that evaluate soil organic carbon (SOC) as a proxy for soil health, combined with county-level data on drought, wildfire, and cropland conditions. You provide data-driven insights to inform policies that promote sustainable land use and agricultural practices. \\ \hline
\end{tabular}
\caption{System Prompts for Agronomist, Farm Consultant, and Policy Maker Roles. Note these prompts are GPT-generated.}
\label{tab:system_prompts}
\end{table}

\clearpage

\section{Data Source Details}
\begin{figure}[h]
  \centering
  \includegraphics[width=1.1\linewidth]{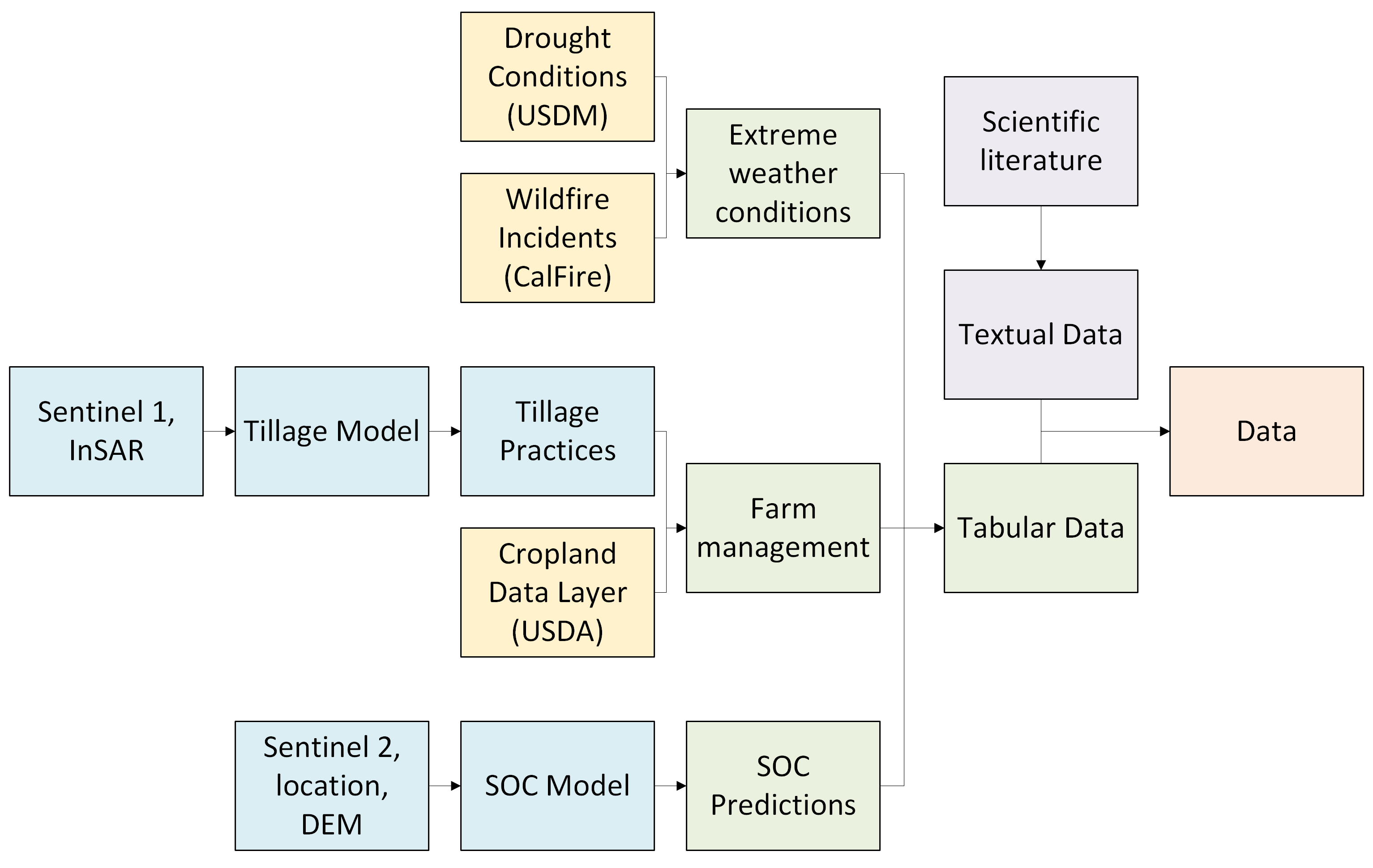}
  \caption{The textual data includes recent scientific literature on how different factors influence SOC. The remaining data is tabular, including extreme weather conditions, farm management practices, and SOC predictions. The majority of the data is public, where the tillage practice predictions and SOC predictions are inferred from machine learning models. The tillage model's inputs include Sentinel 1 satellite imagery and Interferometric Synthetic Aperture Radar (InSAR). The SOC model's inputs include Sentinel 2 satellite imagery, location, and digital elevation model (DEM).}
  \label{fig:copilot_detailed_arch}
\end{figure}

\clearpage
\section{Detailed Results}
\begin{table}[h!]
\centering
\small
\begin{tabular}{p{15cm}}
\multicolumn{1}{c}{\textbf{Large-scale Analysis of Droughts and SOC: San Joaquin County }} \\ \hline
\textbf{Question:} Query the drought conditions and tell me how they influence SOC change in San Joaquin County. Does this align with scientific research? Please provide citations.  \\ \hline
\textbf{SOC Copilot Response:} \\
\begin{minipage}[t]{15cm}
"\#\#\# Drought Conditions in San Joaquin County
San Joaquin County has experienced varying drought conditions over the past decade:
- **2013 - 2016**: Severe drought (D3)
- **2020**: Moderate drought (D1)
- **2021**: Drought (D2)
- **2022**: Severe drought (D3)
These periods coincide with declining soil moisture and varied agricultural strain.
\textcolor{blue}{\#\#\# Influence of Drought on Soil Organic Carbon (SOC) in San Joaquin County
The Soil Organic Carbon (SOC) in San Joaquin County has shown a noticeable decrease from 2016 to 2023:
- **2016**: 3.886\% SOC
- **2023**: 2.644\% SOC}
This decline in SOC could be influenced by several factors including drought conditions reducing soil moisture essential for carbon storage and biological activity.
\#\#\# Scientific Alignment
Research indicates that drought conditions significantly affect SOC levels, primarily through disruption of soil microbial activity and changes in carbon cycling dynamics. Here are the key findings from the scientific literature:
1. **Impact on Microbial Communities**: Drought conditions alter microbial community composition, reducing their ability to process and stabilize soil carbon, thus leading to increased carbon loss ([Bailey et al., Communications Earth \& Environment, 2021](https://doi.org/10.1038/s43247-021-00198-4)). This can be linked with the SOC decrease in San Joaquin, as microbial activity is essential for maintaining SOC levels.
2. **General Findings on Drought and SOC**:
   - Decreased SOC due to reduced moisture inhibits microbial activity and carbon cycling ([Jinquan Li et al., Environmental Research Communications, 2023](https://doi.org/10.1088/2515-7620/ad04f5)).
   - Prolonged drought spells can reduce the SOC pools significantly across different agricultural systems, with management practices playing a crucial role in mitigating these effects.
\#\#\# Summary
The decrease in soil organic carbon in San Joaquin County aligns well with the established scientific understanding that drought adversely affects soil carbon levels by altering microbial activities and disturbing the overall soil carbon cycle. \textcolor{blue}{The regional drought patterns corroborate the scientific literature that suggests increasing drought frequency due to climate change will likely exacerbate these effects, impacting soil health and agricultural sustainability.} Further research and improved agricultural practices are essential for mitigating the adverse effects of these drought conditions on soil carbon stocks."
\end{minipage} \\ \hline

\textbf{GPT-4 Response:} \\
\begin{minipage}[t]{15cm}
"Drought conditions can significantly influence soil organic carbon (SOC) levels, particularly in regions like San Joaquin County, which is part of California's Central Valley. This area is known for its agricultural productivity, which is heavily dependent on both natural soil fertility and irrigation practices.
\#\#\# Influence of Drought on SOC in San Joaquin County
1. **Reduction in Biomass Production**:
   - Drought conditions lead to lower water availability for crops, reducing plant growth and the subsequent biomass returned to the soil. This reduction in organic matter input can decrease the SOC accumulation.
2. **Increased Soil Erosion**:
   - With less vegetation cover during drought periods, soils are more prone to erosion by wind and water. This erosion can lead to a loss of the topsoil, which is rich in organic carbon, thereby reducing the overall SOC stock.
3. **Changes in Microbial Activity**:
   - Soil moisture is a critical factor influencing microbial decomposition of organic matter. During drought, reduced soil moisture can slow down microbial activity, initially leading to a slower SOC turnover. However, prolonged drought can alter microbial community composition and function, potentially leading to increased SOC decomposition once normal conditions resume.
4. **Impact on Soil Structure**:
   - Drought can affect soil structure by causing soil particles to compact and harden, which may reduce the infiltration and storage of water and also impede root growth. This can further stress vegetation and impact the cycling and sequestration of carbon in the soil.
\#\#\# Scientific Research Alignment Research supports the notion that drought conditions can significantly impact SOC dynamics. For instance:
- **Schimel, J.P., et al. (2015)** in their study published in *Nature Climate Change*, highlighted that drought leads to a reduction in soil carbon stocks due to decreased plant inputs and increased respiration rates under stress conditions.
- **Deng, L., et al. (2016)** in *Global Change Biology*, found that prolonged drought affects both the quantity and quality of plant-derived carbon inputs to the soil, altering SOC dynamics.
\#\#\# Local Studies and Data For specific data on San Joaquin County, local agricultural extensions or university research departments like the University of California, Davis, might have conducted detailed studies assessing the impact of recent droughts on SOC. These studies would provide localized insights that are more directly applicable to San Joaquin County rather than broader regional or global studies.
\textcolor{blue}{\#\#\# Conclusion
The influence of drought on SOC in San Joaquin County aligns with broader scientific research, which suggests significant impacts due to changes in biomass production, soil erosion, microbial activity, and soil structure. For precise, localized data, consulting studies specific to San Joaquin County or similar agricultural regions within California would provide the most direct insights.}
\#\#\# Citations To access the specific studies mentioned, you would typically search academic databases like Google Scholar, PubMed, or directly access journals like *Nature Climate Change* and *Global Change Biology*. Unfortunately, I cannot fetch real-time data or access external databases directly to provide specific URLs or complete citation details."
\end{minipage} \\ \hline

\end{tabular}
\caption{Comparison of SOC Copilot and GPT-4 Responses for query about the influence of drought on SOC change in San Joaquin County. With access to SOC prediction data, the SOC Copilot provides more localized insights.}
\label{tab:soc_gpt4_responses_drought}
\end{table}

\begin{table}[h!]
\centering
\small
\begin{tabular}{p{15cm}}
\multicolumn{1}{c}{\textbf{Large-scale Analysis of Wildfires and SOC: Merced vs Sonoma County}} \\ \hline
\textbf{Question:} Query the wildfire conditions and tell me how they influence SOC change in Merced vs Sonoma County.  Does this align with scientific research? Please provide citations.   \\ \hline

\textbf{SOC Copilot Response:} \\
\begin{minipage}[t]{15cm}
"\#\#\# Analysis of Wildfire Incidents and Soil Organic Carbon Change

\#\#\#\# Wildfire Data Overview:
- **Merced County**:
  - Multiple wildfire incidents from 2013 to 2024, indicating a persistent wildfire activity over the years.
- **Sonoma County**:
  - Also experienced significant wildfire activity, with some major fires including the Kincade Fire in 2019.

\textcolor{blue}{\#\#\#\# Soil Organic Carbon (SOC) Changes:
- **Merced County**:
  - SOC 2016: 2.85%
  - SOC 2023: 2.61%
  - There was a **decrease** in SOC over these years.
- **Sonoma County**:
  - SOC 2016: 1.79%
  - SOC 2023: 2.06%
  - There was an **increase** in SOC over these years.}

\#\#\# Scientific Research Alignment:
Research suggests wildfires have significant impacts on soil organic carbon and overall soil health. Key findings from the literature:
1. **Loss of SOC and Nutrients**:
   - Wildfires can lead to a rapid release of soil carbon and nitrogen, thus affecting the SOC levels ([source](https://www.elsevier.com/locate/catena)).
   - Recurrent wildfires particularly degrade soil organic matter and can alter soil properties like texture, cation exchange capacity, and nitrogen content ([source](https://www.elsevier.com/locate/jenvman)).

2. **Vegetation and Recovery**:
   - Post-fire environments see a change in vegetation types which can benefit from the alteration in soil properties ([Catena221(2023)106797](https://www.elsevier.com/locate/catena)).
   - This change in vegetation, in turn, influences subsequent soil organic carbon dynamics, promoting either recovery or further degradation depending on the severity and recurrence of fires.

\#\#\# Discussion
- The differences in SOC trends between Merced and Sonoma County could relate to several factors:
 \textcolor{blue}{ 1. **Severity and Frequency of Fires**: More frequent and severe fires, as seen in Sonoma from 2017 onwards, might have initially degraded SOC but later contributed to an increase due to changes in vegetation and possibly more aggressive soil management and recovery efforts.}
  2. **Management Practices**: Post-fire management practices such as replanting, mulching, or other restorative measures could enhance SOC.
  3. **Type of Vegetation and Soil**: Different soil types and vegetation in Merced and Sonoma might respond differently to wildfires, affecting SOC dynamics.

\#\#\# Conclusion
The observed changes in SOC in Merced and Sonoma align with scientific research indicating that wildfires markedly affect SOC levels. Variations in these effects can depend on local environmental conditions, wildfire management, and recovery practices. For detailed strategies to mitigate these impacts, further localized studies would be beneficial, especially to tailor recovery practices that optimize SOC restoration and preservation post-wildfires. 

\#\#\# Citations:
1. Xiaoying Li et al., ""Impact of wildfire on soil carbon and nitrogen storage..."", Catena, 2023.
2. L. Facudade et al., ""Impact of wildfire recurrence on soil properties and organic carbon fractions"", Journal of Environmental Management, 2024."

\end{minipage} \\ \hline

\textbf{GPT-4 Response:} \\
\begin{minipage}[t]{15cm}
\textcolor{blue}{I currently don't have the capability to directly query real-time or specific historical data, including wildfire conditions or their direct impact on soil organic carbon (SOC) changes in specific locations like Merced or Sonoma County. However, I can provide a general overview based on scientific understanding and research.}

\#\#\# Wildfire Impact on SOC
Wildfires can significantly alter soil organic carbon (SOC) through combustion of organic matter and changes in soil properties. The extent and nature of these changes can depend on the intensity and frequency of the wildfires, as well as the type of vegetation and soil properties.

1. **Combustion of Organic Matter**: High-intensity fires can combust a large portion of aboveground and belowground biomass, directly reducing the organic carbon stock. This combustion releases CO2 into the atmosphere and reduces the input of organic materials into the soil.

2. **Soil Heating**: Wildfires can heat soils to varying degrees depending on fire intensity. Severe heating can lead to the loss of SOC through volatilization and altered microbial activity, which decomposes SOC more rapidly.

3. **Changes in Soil Properties**: Fire can alter soil hydrophobicity, structure, and nutrient availability, which in turn can affect SOC stabilization and formation post-fire.

\#\#\# Comparison Between Merced and Sonoma County
The impact of wildfires on SOC in Merced versus Sonoma County would likely differ based on vegetation type, soil type, climate, and fire history:

- **Vegetation Type**: Sonoma County, with more forested areas, might experience different SOC dynamics compared to the more agricultural and grassland areas of Merced County.
- **Climate and Soil Type**: Differences in climate and soil types between the two counties can influence how SOC responds to wildfires. For instance, wetter conditions in Sonoma might lead to different post-fire SOC dynamics compared to the drier conditions in Merced.

\#\#\# Alignment with Scientific Research
Research indicates that wildfires generally lead to immediate reductions in SOC but can also lead to longer-term changes in SOC dynamics due to alterations in vegetation, soil properties, and microbial activity. The specific impacts and recovery can vary widely based on local conditions and fire characteristics.

\#\#\# Citations
For a deeper understanding, you might refer to these sources:
- Certini, G. (2005). Effects of fire on properties of forest soils: A review. **Oecologia**, 143(1), 1-10.
- DeBano, L. F. (2000). The role of fire and soil heating on water repellency in wildland environments: a review. **Journal of Hydrology**, 231, 195-206.
- Knicker, H. (2007). How does fire affect the nature and stability of soil organic nitrogen and carbon? A review. **Biogeochemistry**, 85(1), 91-118.
For specific data and analysis on Merced and Sonoma County, consulting local environmental agencies or academic institutions conducting regional studies would be necessary. They might provide detailed reports or data on the impact of wildfires on SOC in these specific counties."

\end{minipage} \\ \hline

\end{tabular}
\caption{Comparison of SOC Copilot and GPT-4 Responses for query about the influence of wildfires on SOC change in Merced vs Sonoma County. With access to SOC prediction data, the SOC Copilot provides more localized insights.}
\label{tab:soc_gpt4_responses_wildfire}
\end{table}

\begin{table}[h!]
\centering
\small
\begin{tabular}{p{15cm}}
\multicolumn{1}{c}{\textbf{Effect of Farm Management on SOC: Tillage}} \\ \hline
\textbf{Question:} How do the tillage practices in farms in Monterey vs Tulare County influence SOC change in conjunction with factors including drought, wildfire, and crops? \\ \hline

\textbf{SOC Copilot Response:} \\
\begin{minipage}[t]{15cm}
"\#\#\# Comparative Analysis of Tillage Practices in Monterey and Tulare Counties:
\#\#\#\# Tillage Practices:
\textcolor{blue}{- **Monterey County**: No tillage practice (value: 0.0), indicating an approach that preserves soil structure and moisture.
- **Tulare County**: High-intensity tillage practices (value: 1.0), which tend to disrupt soil structure and can lead to loss of soil organic carbon.}
\textcolor{blue}{\#\#\#\# Changes in Soil Organic Carbon (SOC):
- **Monterey County**: 
  - SOC 2016: 2.39%
  - SOC 2023: 2.00%
  - Decrease in SOC, possibly due to other environmental factors as no tillage was practiced here.
- **Tulare County**:
  - SOC 2016: 5.58%
  - SOC 2023: 5.48%
  - Slight decrease in SOC despite high-intensity tillage, which usually is expected to reduce SOC due to increased oxidation and erosion.}
\#\#\#\# Drought Conditions Impact:- Both counties exhibit similar drought histories with severe drought conditions in 2014-2016 and 2021, with slight improvements in non-drought years. Drought can stress plants, reduce biomass production, and subsequently, SOC when plant residue input into the soil reduces.
\#\#\#\# Crop Impact:- **Monterey County**: Variety of crops with a focus on high-residue crops like broccoli which might help in maintaining some level of SOC due to residual biomass.
- **Tulare County**: Diverse agricultural production including high residue crops like broccoli and others like grains and tree crops which influence SOC dynamics differently. Tree crops like almonds may contribute to higher SOC due to deeper rooting systems and longer-term soil structure stability.
\#\#\# Discussion:The practice of high-intensity tillage in Tulare versus no-tillage in Monterey shows differing impacts on soil health and SOC levels. High-intensity tillage generally leads to quicker SOC loss due to increased exposure to erosion and oxidation, yet the drop in SOC in Tulare is surprisingly lower than expected. This indicates other factors such as crop choice (type and diversity), soil management practices in response to drought, and perhaps adoption of other soil conservation measures after tillage can ameliorate some negative effects of tillage on SOC.
Monterey's no-tillage approach generally helps in preserving SOC but the still observed decrease could be influenced by prolonged drought impacts which reduce biomass production, and hence, organic carbon returns to the soil.
\#\#\# Conclusion: Tillage practices, along with drought and crop types, significantly influence SOC trends in both counties. \textcolor{blue}{Despite high-intensity tillage, Tulare has managed to maintain relatively high SOC levels, likely through integrated soil health practices and diverse cropping systems contributing positively to soil organic matter. Conversely, despite no tillage, the SOC in Monterey has decreased, likely influenced by severe external stresses like drought impacting biomass production negatively.}
Both counties demonstrate the critical balance in agricultural management practices (tillage, drought response, and crop choices) to maintain and enhance soil organic carbon, vital for soil fertility and reducing atmospheric CO2 levels."
\end{minipage} \\ \hline
\textbf{GPT-4 Response:} \\
\begin{minipage}[t]{15cm}
"The influence of tillage practices on soil organic carbon (SOC) change in Monterey and Tulare Counties, especially in the context of external factors like drought, wildfire, and crop types, is a complex interplay of agricultural management, environmental conditions, and land use history. Here’s a detailed exploration:
\textcolor{blue}{\#\#\# 1. **Tillage Practices**
   - **Monterey County**: Known for its diverse agriculture, including vegetables, fruits, and vineyards, Monterey typically employs varied tillage practices. Conservation tillage is increasingly popular in vegetable production to reduce erosion and improve water retention.
   - **Tulare County**: Dominated by intensive agriculture, especially dairy, nuts, and fruits, Tulare often uses more traditional tillage practices. However, there is a growing shift towards reduced tillage to conserve water and improve soil structure.}
\#\#\# 2. **Soil Organic Carbon (SOC) Dynamics**
   - **Impact of Tillage**: Traditional tillage accelerates the decomposition of organic matter, releasing CO2 into the atmosphere, thus reducing SOC. Conservation or reduced tillage, on the other hand, helps in the accumulation of SOC by minimizing soil disturbance and enhancing soil structure.
   - **Monterey vs. Tulare**: The extent of SOC change due to tillage might be more pronounced in Tulare due to its heavier reliance on traditional tillage methods compared to Monterey, where conservation practices are more common.
\#\#\# 3. **Influence of Drought**
   - **Soil Moisture and SOC**: Drought conditions, prevalent in both counties, reduce soil moisture, impacting microbial activity and thus SOC dynamics. Reduced microbial activity slows down the decomposition of organic matter.
   - **Adaptation Practices**: In response to drought, farmers might adopt more conservation tillage to retain soil moisture, indirectly promoting SOC sequestration.
\#\#\# 4. **Impact of Wildfires**
   - **Immediate Effect**: Wildfires can lead to a significant loss of SOC due to combustion of organic matter. Post-fire erosion can further deplete SOC.
   - **Long-term Impact**: The regrowth phase after a wildfire can potentially increase SOC if managed properly, depending on the recovery of vegetation and subsequent land management practices.
\#\#\# 5. **Influence of Crop Types**
   - **Monterey’s Crops**: The diversity of crops, including high-residue crops like broccoli or lettuce, can contribute positively to SOC through the addition of organic matter from crop residues.
   - **Tulare’s Crops**: Permanent crops like orchards might contribute differently to SOC dynamics. The continuous root systems and lesser soil disturbance (in cases of reduced tillage) can enhance SOC.
\#\#\# 6. **Comparative Analysis**
   - **Management Practices**: The choice of crop and tillage practice in response to environmental pressures like drought and wildfire can significantly influence SOC. Monterey’s shift towards conservation practices might be enhancing SOC more effectively than Tulare’s more traditional approaches.
   - **Environmental and Climatic Factors**: Both counties face environmental stresses that necessitate adaptive management strategies. The effectiveness of these strategies in mitigating SOC loss or promoting SOC sequestration can vary based on how aggressively and swiftly they are implemented.
\#\#\# 7. **Future Considerations**
   - **Research and Monitoring**: Continuous monitoring of SOC changes, coupled with research into the impacts of different tillage and crop management practices under changing climatic conditions, is crucial.
   - **Policy and Guidance**: Both counties could benefit from policies that promote sustainable practices, particularly those that enhance SOC sequestration, such as incentives for reduced tillage and the use of cover crops.
\textcolor{blue}{In conclusion, the tillage practices in Monterey and Tulare Counties, along with the influence of drought, wildfire, and crop types, play significant roles in determining the dynamics of SOC.} Adaptive management strategies that consider local environmental conditions and global climate change impacts are essential for sustainable agricultural practices and SOC management."
\end{minipage} \\ \hline

\end{tabular}
\caption{Comparison of SOC Copilot and GPT-4 Responses for query about the influence of tillage on SOC change in Monterey vs Tulare County. While GPT-4 makes broad analysis, the SOC Copilot provides more nuanced insights based on tillage detection data.}
\label{tab:soc_gpt4_responses_tillage}
\end{table}

\begin{table}[h!]
\centering
\small
\begin{tabular}{p{15cm}}
\multicolumn{1}{c}{\textbf{Effect of Farm Management on SOC: Regenerative Practice Comparison}} \\ \hline
\textbf{Question:} Gable Farms in Riverside County implemented planting while Doug and Cathy Ielmorini Dairy in Marin County implemented the regenerative practice of composting in 2021-22. How do the regenerative practices among other factors like drought and wildfire influence the change in SOC? \\ \hline
\textbf{SOC Copilot Response:} \\
\begin{minipage}[t]{15cm}
"\#\#\# Analysis of Regenerative Practices Influences on SOC in Riverside and Marin Counties:
\#\#\#\# Regenerative Practices:
- **Riverside County (Gable Farms)**: Implementation of planting practices which are generally aimed to enhance biomass production and consequently soil organic carbon (SOC) through root growth and residual organic matter.
- **Marin County (Doug and Cathy Ielmorini Dairy)**: Use of composting which can directly contribute to increased soil organic matter, SOC content, and overall soil fertility.
\textcolor{blue}{\#\#\#\# SOC Changes:
- **Riverside County**:
  - SOC 2016: 2.99%
  - SOC 2023: 0.94%
  - A significant reduction in SOC despite regenerative planting, possibly due to other factors like drought and frequent wildfires which can contribute to organic matter loss.
- **Marin County**:
  - SOC 2016: 1.96%
  - SOC 2023: 1.92%
  - Slightly decreased SOC, which might suggest that composting helped in buffering the SOC loss considering the drought and wildfire pressures.}
\#\#\#\# Drought Conditions:
- Both counties have experienced severe to moderate drought conditions over several years which can severely limit plant growth and hence reduce soil organic carbon inputs from plant residues.
\#\#\#\# Wildfire Impact:
- **Riverside County**: Experienced a large number of wildfire incidents over the decade which would have directly impacted organic matter through combustion and indirectly by affecting soil properties that support SOC retention.
- **Marin County**: While fewer in number compared to Riverside, the wildfires still present a significant disturbance which can disrupt soil layers and lead to SOC loss.
\#\#\# Discussion:
\textcolor{blue}{The decline in SOC in Riverside County could predominantly be influenced by the intense and frequent wildfires along with severe drought, which despite any benefits from planting regenerative practices, likely had an overwhelming negative impact on SOC levels. In contrast, composting practices in Marin County may have provided some protection against SOC loss through direct inputs of organic matter and enhancing soil moisture retention during drought conditions—which helps in reducing the impact of drought and wildfires on SOC.
\#\#\# Conclusion:
The results suggest that while regenerative practices like planting and composting can support soil carbon sequestration, their effectiveness can be heavily moderated by external pressures such as drought and wildfires.} This is evident from the stark difference in SOC trends between Riverside and Marin given similar drought histories but different intensities and frequencies of wildfires and regenerative strategies. Integration of regenerative practices in areas prone to frequent natural disturbances must be accompanied by other protective measures such as fire breaks, irrigation enhancements during droughts, and diversified plantings to buffer against the diverse impacts on SOC.
The comparison underlines the importance of a holistic approach to land management that considers all potential environmental stressors to optimize the benefits of regenerative agricultural practices in supporting soil health and carbon sequestration."
\end{minipage} \\ \hline
\textbf{GPT-4 Response:} \\
\begin{minipage}[t]{15cm}
"The implementation of regenerative agricultural practices such as planting and composting, as seen at Gable Farms and Doug and Cathy Ielmorini Dairy, can significantly influence soil organic carbon (SOC) levels, which are crucial for soil health, fertility, and overall ecosystem functioning. Here’s a detailed look at how these practices, along with external factors like drought and wildfire, might affect SOC:
\#\#\# 1. **Regenerative Practices and SOC**
\#\#\#\# **Planting**
- **Increased Biomass**: Planting more crops or cover crops increases the amount of biomass returned to the soil. This biomass eventually decomposes and contributes to the SOC pool.
- **Carbon Sequestration**: Plants absorb CO2 from the atmosphere during photosynthesis, some of which is transferred to the soil and stored as SOC.
\#\#\#\# **Composting**
- **Nutrient Enrichment**: Composting adds organic matter rich in nutrients to the soil, which can enhance microbial activity and accelerate the formation of humus, a stable form of SOC.
- **Reduction in Soil Erosion**: Enhanced soil structure reduces erosion, thereby preventing the loss of topsoil and SOC.
\#\#\# 2. **Drought and SOC**
- **Reduced Plant Growth**: Drought conditions lead to reduced biomass production as plants struggle to grow without adequate water, potentially decreasing the amount of organic matter added to the soil.
- **Microbial Activity**: Drought can suppress microbial activity necessary for the decomposition of organic matter into SOC.
\#\#\# 3. **Wildfire and SOC**
- **Immediate Loss of SOC**: High-intensity wildfires can combust organic matter in the soil, leading to an immediate loss of SOC.
- **Long-term Changes**: Post-fire, the soil can either gain SOC through the incorporation of ash and charred material or lose SOC due to increased erosion if vegetation recovery is slow.
- **Altered Microbial Communities**: Fire can alter the composition and function of soil microbial communities, affecting the rates of SOC decomposition and formation.
\#\#\# 4. **Interactions and Overall Impact**
The overall impact on SOC from these practices and environmental factors is complex and influenced by interactions among them. For example:
- **Composting might help mitigate the negative effects of drought** on SOC by improving soil moisture retention.
- **Planting after a wildfire** can help stabilize the soil and add organic matter, aiding in the recovery of SOC levels.
\#\#\# 5. **Monitoring and Management**
To effectively manage SOC levels, continuous monitoring of soil health indicators is essential, especially following the implementation of regenerative practices or after events like droughts or wildfires. Adjustments in management practices might be necessary based on observed changes in SOC and overall soil health.
\textcolor{blue}{\#\#\# Conclusion
Regenerative practices like those implemented by Gable Farms and Doug and Cathy Ielmorini Dairy are generally beneficial for increasing SOC levels, which can help improve soil health, enhance productivity, and mitigate climate change. However, the effectiveness of these practices can be moderated by external factors such as drought and wildfire, highlighting the need for adaptive management strategies in the face of environmental changes."}
\end{minipage} \\ \hline
\end{tabular}
\caption{Comparison of SOC Copilot and GPT-4 Responses for query comparing the influence of regenerative practices on SOC change in Riverside vs Marin County. While GPT-4 makes general analysis, the SOC Copilot provides more nuanced insights based SOC predictions and environmental data. The SOC Copilot suggests that environmental conditions have heavier influence on SOC change compared to regenerative practices. In addition, the practice of compost may be more effective than planting in improving soil health.}
\label{tab:soc_gpt4_responses_practice}
\end{table}

\begin{table}[h!]
\centering
\begin{tabular}{p{3cm} p{12cm}}  
\multicolumn{2}{c}{\textbf{Personalized Stakeholder Responses}} \\ \hline
\multicolumn{2}{p{15cm}}{\textbf{Question:} "Gaviota Givings, Orella Ranch in Santa Barbara County implemented the regenerative practice of planting in 2021-22. How does this influence the change in SOC among other factors like drought and wildfire? Based on your role, what specific insights can you draw?"} \\ \hline
\textbf{Role} & \textbf{Response Excerpt} \\ \hline
\textbf{Agronomist} & 
"\#\#\#\# Recommendations:
\textcolor{blue}{1. **Enhanced Water Management:** Implementing more efficient irrigation systems and drought-tolerant crop varieties could mitigate some of the drought-related declines in crop productivity and, indirectly, SOC levels.}
2. **Post-Wildfire Soil Rehabilitation:** Immediate measures to restore soil structure and prevent erosion after fires can help in quicker recovery of SOC levels.
3. **Long-term Monitoring:** Continuous assessment of SOC along with other soil health indicators is essential to understand the full impact of regenerative practices amid changing climatic conditions and frequent natural disturbances."\\ \hline
\textbf{Farm Consultant} & 
"**Insights and Advice:**
1. **Scale and Intensify Regenerative Practice:** Given the stress from drought and wildfires, it might be beneficial to expand regenerative agricultural methods across more areas of the county. Practices such as cover cropping, reduced tillage, and diverse crop rotations can help build a more resilient soil system.
2. **Monitor SOC Progression:** Continuing to monitor SOC changes more closely at the ranch level can provide insight on whether adjustments in the regenerative practices are needed or if these practices are leading to desired outcomes.
\textcolor{blue}{3. **Integrate Additional Resilience Practices:** Given the recurrent wildfires, integrating fire resilience strategies, such as creating firebreaks, choosing fire-resistant plant species, or even controlled burns under safe conditions, could help protect the soil and vegetation.
4. **Water Management Innovations:** To combat the effects of prolonged drought periods, implement more efficient water use strategies, such as drip irrigation, and consider the suitability of crops with lower water demands or those adapted to drought conditions."} \\ \hline
\textbf{Policymaker} & 
"\#\#\# Recommendations for Policy and Further Actions:
1. **Persistent Monitoring and Support**: Encourage and support continued and enhanced regenerative practices at Orella Ranch and similar farms across the county, focusing on long-term improvement and stability of SOC.
2. **Comprehensive Management Strategies**: Develop and implement integrated land management strategies that consider drought management, fire risk reduction, and enhanced regenerative practices to optimize soil health and productivity.
3. \textcolor{blue}{**Research and Data Analysis**: Foster research initiatives that systematically analyze the long-term impacts of regenerative practices on SOC under varying climatic stress conditions to refine agricultural policy making and support.}
In summary, while the immediate changes in SOC might not fully reflect the benefits of the regenerative practices due to confounding environmental challenges, the comprehensive approach provided by these practices holds potential for more significant positive shifts in soil health over a prolonged period. \textcolor{blue}{Implementing proactive and supportive policies will be crucial in realizing their full potential."} \\ \hline
\end{tabular}
\caption{Response Excerpts for Agronomist, Farm Consultant, and Policy Maker Roles for the query about the impact of regenerative practices on SOC change at Orella Ranch, Santa Barbara County. This exemplifies the SOC Copilot's capability to provide role-specific insights to different stakeholders.}
\label{tab:persona_responses}
\end{table}

\end{document}